\documentstyle[prl,aps,graphicx,multicol]{revtex}
\newcommand{\ket}[1]{\left| #1 \right\rangle}

\newcommand{\braket}[1]{\left\langle #1 \right\rangle}

\begin{document}

\title{
Charge Echo in a Cooper-Pair Box
}
\author{
Y. Nakamura,$^1$ Yu.\ A. Pashkin,$^2$ T. Yamamoto,$^1$ and J. S. Tsai$^1$
}
\address{
$^1$NEC Fundamental Research Laboratories, Tsukuba, Ibaraki 305-8501, Japan\\
$^2$CREST, Japan Science and Technology Corporation (JST), Kawaguchi, Saitama, 332-0012, Japan
}
\date{\today}
\maketitle

\begin{abstract}
A spin-echo-type technique is applied to an artificial two-level system that utilizes charge degree of freedom in a small superconducting electrode.
Gate-voltage pulses are used to produce the necessary pulse sequence in order to eliminate the inhomogeneity effect in the time-ensemble measurement and to obtain refocused echo signals.
Comparison of the decay time of the observed echo signal with estimated decoherence time suggests that low-frequency energy-level fluctuations due to the $1/f$ charge noise dominate the dephasing in the system.
\end{abstract}
\pacs{PACS numbers: }

\begin{multicols}{2}


In a small-Josephson-junction circuit, we can construct an artificial two-level system, which is expected to have a long decoherence time and work as a qubit \cite{Makhlin01}.
There are two types of qubit, depending on which, either charge or phase, degree of freedom is used in the two-level system.
The charge qubit, a Cooper-pair box, uses two charge states in a small superconducting electrode connected to a reservoir via a Josephson junction \cite{Shnirman97}, while the phase qubit uses two phase states in a small superconducting loop intersected by Josephson junction(s) \cite{Mooij99}.
Superposition of the two charge states \cite{Bouchiat98,Nakamura97} and that of two phase states \cite{Friedman00,vanderWal00} have been observed recently, and coherent control of the charge qubit has also been demonstrated \cite{Nakamura99}.
Being solid-state devices, these qubits are also expected to be scalable to an integrated system for quantum information processing \cite{Nielsen00}. 
However, the actual decoherence time and the decoherence mechanism are not yet fully understood.
Accordingly, we performed an experiment on the decoherence in a Cooper-pair box.

In this experiment, a Cooper-pair box is subjected to a sequence of gate-voltage pulses, which give rise to quantum-state control similar to those in the free-induction-decay (FID) and spin-echo experiments in the field of nuclear magnetic resonance \cite{Slichter90}.
Although our experiment is on a {\it single} two-level system, such technique is useful for studying the effect of fluctuations, that is, temporal inhomogeneities, in the {\it time-ensemble} measurement.
We observed rapid decay of the FID signal and significant recovery of the coherence signal in the echo experiment.
These observations indicate that the effect of low-frequency energy-level fluctuations is dominant in the dephasing of the two-level system.
Furthermore, we compare the decay of the observed ``charge-echo'' signal with estimated decoherence due to several possible sources, and suggest a contribution of $1/f$ background charge noise to the dephasing.


The Cooper-pair-box device \cite{Nakamura99} consists of a small ``box'' electrode coupled to a reservoir electrode via a Josephson junction, two gate electrodes capacitively coupled to the box, and an additional probe electrode connected to the box via a highly resistive tunnel junction (Fig.~1).
All the electrodes are made of Al with a superconducting gap energy $\Delta$ larger than the other important energy scales; thus, quasiparticle excitation can be neglected (except for the measurement process described below).
Because of the charging effect, it is a good approximation of the quantum state of the device to take into account only two charge states with the lowest energies---the ones closest to the charge neutrality of the box---which differ by a single Cooper pair.
For example, if the total gate-induced charge in the box, $Q_t \equiv Q_0+C_p V_p(t)$ ($Q_0 \equiv C_g V_g + C_b V_b$), is close to one, the relevant charge states are $\ket{n=0}$ and $\ket{1}$, where $n$ is the excess number of Cooper pairs in the box and $C_i$ ($i=p,g,b$) is the capacitance between the box and each electrode voltage-biased at $V_i$.

On the basis of $\ket{\uparrow}\equiv \ket{0}$ and $\ket{\downarrow}\equiv \ket{1}$, the effective Hamiltonian of the two-level system is $H=\frac{1}{2} \delta E (Q_t) \sigma_z - \frac{1}{2} E_J \sigma_x$, where $\delta E(Q_t) \equiv 4E_C (Q_t/e-1)$ is the energy difference between the two charge states, $E_C \equiv e^2/(2C_\Sigma)$ is the single-electron charging energy of the box, $C_\Sigma$ is the total capacitance of the box, $E_J$ is the Josephson energy of the junction, and $\sigma_z$ and $\sigma_x$ are Pauli matrices.
Therefore, the quantum-state evolution of the Cooper-pair box is mapped on the evolution of the fictitious spin-$\frac{1}{2}$ state in an effective magnetic field, $\vec{B}=(B_x, 0, B_z)=(E_J, 0, -\delta E(Q_t))$.

In the present device, the energy scales were determined by independent measurements as $\Delta = 235 \pm 10$~$\mu$eV, $E_C=122 \pm 3$~$\mu$eV, and $E_J=34 \pm 2$~$\mu$eV \cite{comment2}.
The present experiment was made at the temperature of about 30--60~mK ($k_B T\simeq 3$--5~$\mu$eV).
The details of the experimental setup have been published elsewhere \cite{Nakamura00}.

Quantum-state control is realized by application on a high-speed gate-voltage pulse $V_p(t)$ to modulate $Q_t$ non\-adiabatically \cite{Nakamura99}.
In terms of fictitious spin, $B_z$ is abruptly changed, and the spin precesses around the direction of the new magnetic field until $B_z$ is switched back after the pulse width $\Delta t$, as depicted in the Bloch spheres in Fig.~1(b).
We set our working point at $Q_0=0.45e$ where $|\delta E(Q_0)| \simeq 270$~$\mu$eV $\gg E_J$.
A $\pi$-pulse is produced by a pulse height $\Delta Q_p=0.55e$ which makes the two states degenerate (i.e., $B_z =0$) during the pulse (Fig.~1(b) top).
We found that a nominal pulse width $\Delta t=80$~ps is suitable for the $\pi$ rotation.
For a $\frac{\pi}{2}$-pulse, we could, in principle, use a pulse with the same $\Delta Q_p$  and with a different $\Delta t$.
However, for convenience, instead we use a pulse with the same $\Delta t$ and a slightly detuned $\Delta Q_p$ ($B_z>0$).
The bottom sphere in Figure~1(b) illustrates the corresponding schematic quantum-state evolution.
Though the pulse is not equivalent, {\it e.g.}, to a $\left( \frac{\pi}{2} \right)_x$-pulse (which rotates the spin around $x$-axis), it creates fifty-fifty superposition of $\ket{0}$ and $\ket{1}$ and is thus sufficient for our purpose.
For simplicity, in the following argument, we use ideal $(\pi)_x$- and $(\frac{\pi}{2})_x$-pulses and neglect the deviations in the real operations, because such deviations only cause quantitative difference in the absolute signal size but not in the relative decay of the signal \cite{comment3}.
The delay time between the pulses is controlled by programmed patterns in a multi-channel pulse-pattern generator and the shift in a mechanical delay line.

The quantum state after the control is measured by using a probe junction.
The probe junction is voltage-biased appropriately (at $V_b=645$~$\mu$V) so that only the $\ket{1}$ state allows two electrons (quasiparticles) to escape from the box at rates $\Gamma_{\rm qp1}$ and $\Gamma_{\rm qp2}$, while the $\ket{0}$ state does not.
The resistance of the probe junction is made to be very large ($R_b = 42.6$~M$\Omega$) in order to suppress Josephson coupling and to avoid too much decoherence due to the read-out.
The measurement basis is $\sigma_z$-basis, and the quantum state is initialized very closely to $\ket{0}$ after the read-out.
On the other hand, since the output of the single read-out is just two electrons, it is necessary to repeat the read-out process many times on identically prepared quantum states in order to obtain a detectable dc current proportional to the population of $\ket{1}$, i.e., $\braket{\sigma_z}+1$.
The repetition time $T_r$ (128~ns unless specially mentioned) is taken to be long enough ($T_r \gg 1/\Gamma_{\rm qp1}+1/\Gamma_{\rm qp2}$) for efficient measurement and good initialization for the next pulse operation but short enough for a sufficient current signal (at most $2e/T_r$).
The probe current is measured with a time window of 20~ms.
Therefore, each data point is a result of averaging over a time-ensemble of about $10^5$.


We first measured the free-induction decay (FID) by using two $(\frac{\pi}{2})_x$-pulses with a delay time $\delta t_3$.
The first pulse prepares a superposition, and the quantum state acquires phase $\varphi=\Delta E(Q_0) \delta t_3 /\hbar $ during the delay time, where $\Delta E(Q_0)=\sqrt{\delta E(Q_0)^2 +{E_J}^2}$ is the energy difference between the two eigenstates and corresponds to the strength of the fictitious field.
The second pulse projects the phase information on the measurement basis, i.e., $\sigma_z$-basis.
As shown in Fig.~2(c) the signal oscillates with a period of about 15~ps, which agrees well with $h/\Delta E(Q_0)$.
Moreover, it is clear that the FID signal decays quite rapidly within a few hundred picoseconds.
Compared with the previous result in which coherent oscillations between two degenerate states were observed up to about 2~ns \cite{Nakamura99,comment1}, the FID time scale is much shorter.
Since $\frac{d(\Delta E)}{dQ_0}$ equals zero at the degeneracy point but not at the present working point and the inelastic relaxation rate of the two-level system is estimated to be much longer (see below), the short FID time can be attributed to the dephasing due to charge fluctuations.
Our measurement is an average over a time-ensemble of 20~ms, so even very low-frequency fluctuations can contribute, as inhomogeneities in the ensemble, to the fast decay time in the FID experiment.

To remove such an effect, the three pulses shown in Fig.~2(b) are used for the echo experiment.
The second pulse flips the possibly dephased ensemble of spins after the first delay time $t_d$, and during the second delay time, provided that the correlation time of the fluctuations in $\Delta E(Q_0)$ is much longer than $t_d$, the spins respectively precess the same amount as before and are refocused at the end.
The third pulse is again needed for the projection of the phase information onto $\braket{\sigma_z}$.

The observed echo signal is shown in Fig.~2(d).
When the position of the third pulse $\delta t_3$ was swept, an oscillating signal was observed only in the vicinity of $\delta t_3=0$ and decayed as $|\delta t_3|$ increases within nearly the same time scale as the FID signal did.

Decay of the echo signal as a function of the delay time $t_d$ is shown in Fig.~2(f).
The echo amplitude $\Delta I$ was obtained by fitting the echo signal as a function of $\delta t_2$ with a sinusoidal curve (Fig.~2(e)).
The period of the echo signal is about half of those in Figs.~2(c) and (d), because here the difference between the first and the second delay times $(t_d+\delta t_2)-(t_d-\delta t_2)=2\delta t_2$ matters in the final phase.
The slight phase offset at $\delta t_2=0$ is due to an offset in the delay line.
In Fig.~2(f), the decay time is much longer than that of the FID signal.
This implies that dephasing in the FID is mainly due to the low-frequency ($<1/t_d$) part of the fluctuations which is greatly cancelled out by the echo technique.

The above results lead to the question concerning the origin of the decay of the echo signal.
In order to compare with theoretical estimations of a decoherence factor $\braket{\exp i \varphi}$, the echo amplitude data (Fig.~2(f)) and the FID data (Fig.~2(c)) are normalized by fitting with a gaussian curve and are plotted in Fig.~3 on a logarithmic scale.

The well-known $1/f$-like background charge noise, which is believed to be caused by charges randomly fluctuating around the device \cite{Zimmerli92,Zorin96}, may contribute to the decoherence.
The $1/f$ noise measurement is usually restricted to a very low frequency below about 1~kHz, and the noise spectrum in the high-frequency range is unknown.
Nevertheless, here we assume that the $1/f$ spectrum extends to the infinite frequency \cite{Covington00}.
Generally, the dephasing factor in the presence of energy-level fluctuation spectrum $S_{\Delta E}(\omega)$, under the assumption of gaussian fluctuations, is given as 
\begin{eqnarray}
\braket{\exp i\varphi (\tau)} = \exp \left[ -\frac{1}{2} \braket{ \left( \frac{1}{\hbar} \int_{0}^{\tau} dt\, \widetilde{\Delta E} \right)^2} \right] \nonumber \\
=\exp \left[ -\frac{1}{2\hbar^2} \int_{\omega_m}^{\infty} d\omega\, S_{\Delta E} (\omega) \left(\frac{\sin \left( \omega \tau /2 \right)}{\omega /2} \right) ^2 \right]
\end{eqnarray}
for the FID case.
Here $\widetilde{\Delta E}$ is the deviation of $\Delta E$ from its average and  $\omega_m\equiv 2\pi/t_m$ is the low-frequency cutoff due to the finite data-acquisition time $t_m$ (20~ms in the present work).
Similarly, $\braket{\exp i\varphi}$ in the echo experiment becomes
\begin{equation}
\exp \left[ -\frac{1}{2\hbar^2} \int_{\omega_m}^{\infty} d\omega\, S_{\Delta E} (\omega) \left(\frac{\sin ^2 \left( \omega \tau /4 \right)}{\omega /4} \right) ^2 \right]
\end{equation}
because of the cancellation of the dephasing before and after the $\pi$-pulse.
With a $1/f$ spectrum, $S_{\Delta E}(\omega)\simeq \left( \frac{4E_C}{e} \right)^2 S_e(\omega)\equiv \left( \frac{4E_C}{e} \right) ^2 \frac{\alpha}{\omega}$, both Eqs.(1) and (2) give gaussian-type decay \cite{Cottet00}.
For $\alpha=(1.3\times 10^{-3}e)^2$, which we determined by a standard noise measurement on the present device used as a single-electron transistor, the two curves are plotted in Fig.~3, as well as another curve for echo experiment with $\alpha=(0.3\times 10^{-3}e)^2$, which is a typical number given in the literature.
Considering the uncertainty in the noise spectrum, it seems that the curves reasonably reproduce the significant increase of the decay time in the echo signal and the time scale of the gaussian decay qualitatively, suggesting that the $1/f$ noise is the main dephasing source.
It is worth mentioning that, if this is true, long decoherence time required for quantum computing is only be possible after a drastic reduction of the $1/f$ noise.
Because of the long tail to the high frequency, echo technique is not a perfect solution for suppressing the dephasing \cite{Viola98}.

Dephasing due to the electromagnetic environment \cite{Makhlin01} is also calculated by Eq.(1) with $S_{\Delta E}(\omega)=4e^2 \kappa_i^2 S_{Vi}(\omega)$, where $S_{Vi}(\omega) \equiv \frac{\hbar \omega}{2\pi} 2 R_{\rm env} \coth \! \left( \frac{\hbar \omega}{2 k_BT_{\rm env}} \right) $ is the voltage fluctuation (Johnson-Nyquist noise) on each electrode, $\kappa_i\equiv C_i/C_\Sigma$ is the coupling constant, and $T_{\rm env}$ and $R_{\rm env}$ are the temperature and the impedance real part of the environment, which is assumed to be ohmic.
In the relevant temperature range, the dephasing gives exponential decay of $\braket{\exp i\varphi}$ at the rate of $8\pi \frac{k_B T_{\rm env}}{\hbar} \kappa_i^2 \frac{R_{\rm env}}{R_K}$ ($R_K\equiv \hbar/e^2$).
By assuming $T_{\rm env}$ and $R_{\rm env}$ for the pulse gate, which is connected to the pulse generator via a low-loss line, as 300~K and 50~$\Omega$, and for the dc gate and the probe electrode, which are heavily filtered, as 100~mK and 100~$\Omega$ ($10^3 \kappa_i\simeq 1.3$, 2.5, and 73 for $i=p,g,b$), we evaluated the total dephasing rate as (100~ns)$^{-1}$.
Note that the echo technique does not reduce the dephasing caused by white noise.
Also, inelastic relaxation rate due to the environment is estimated from the standard perturbation theory as $\frac{2\pi}{\hbar^2} e^2 \kappa_i^2 \left( \frac{E_J}{\Delta E} \right)^2 S_{Vi} \left (\frac{\Delta E}{\hbar} \right) $.
The total relaxation rate calculated by using the above parameters yields (580~ns)$^{-1}$.
The total decay of the coherence due to the environment (dotted line in Fig.~3), including the dephasing and the relaxation, is much weaker than the decay of the observed echo signal.
It should be mentioned, however, that in the present experiment, the contribution of residual noise in $V_b$ to the observed dephasing cannot be thoroughly excluded.
In particular, the relatively short decay times of FID signal implies the existence of excess noise in the low-frequency range.

The coherence is also disturbed by the read-out process at the probe junction.
The first electron tunnels out from the box, i.e., the system escapes from the relevant Hilbert space of the two-level system, at the rate $\Gamma_{\rm qp1}=\frac{\Delta E_{\rm qp1}}{e^2 R_b} W \! \left( \frac{\Delta E_{\rm qp1}}{2\Delta} \right)$, where $\Delta E_{\rm qp1}\equiv eV_b +E_C+E_C(1-2Q_0/e)$ is the energy dissipated in the quasiparticle tunneling and $W(x)\sim 1$ is a factor related to the BCS quasiparticle density of states \cite{Barone82}.
In the present device, $\Gamma_{\rm qp1}$ is about (8~ns)$^{-1}$ (dashed line in Fig.~3), which should contribute to the decay of the echo signal, but does not explain it completely.


In conclusion, we have demonstrated charge echo in a Cooper-pair box.
The fact that the echo signal survives much longer than the free-induction decay signal indicates that the dephasing is mainly caused by low-frequency energy-level fluctuations.
The observed dephasing behavior agrees qualitatively with the prediction of that due to $1/f$ background charge noise.


We thank Isaac Chuang and Jonathan Jones for suggesting the echo experiment and Daniel Esteve for valuable discussions on dephasing.
This work has been supported by the CREST project of Japan Science and Technology Corporation (JST).

\begin{figure}
\begin{center}
\includegraphics[width=68mm]{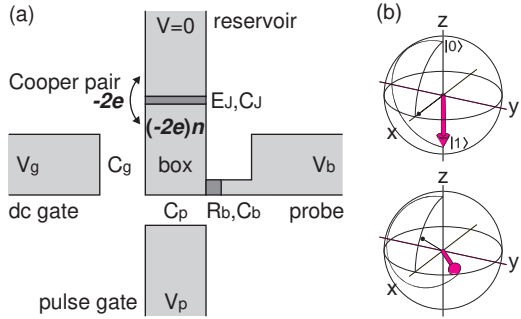}
\end{center}
\caption{
(a) Schematic of a Cooper-pair box with an additional probe electrode.
(b) Bloch sphere representations of schematic quantum-state evolutions at $Q_0=0.45e$ corresponding to the two pulse heights $\Delta Q_p=0.55e$ (top) and $\Delta Q_p=0.53e$ (bottom).
The thin arrow in the $xz$-plane indicates the direction of the effective magnetic field.
}\label{fig:device}
\end{figure}


\begin{figure}
\begin{center}
\includegraphics[width=85mm]{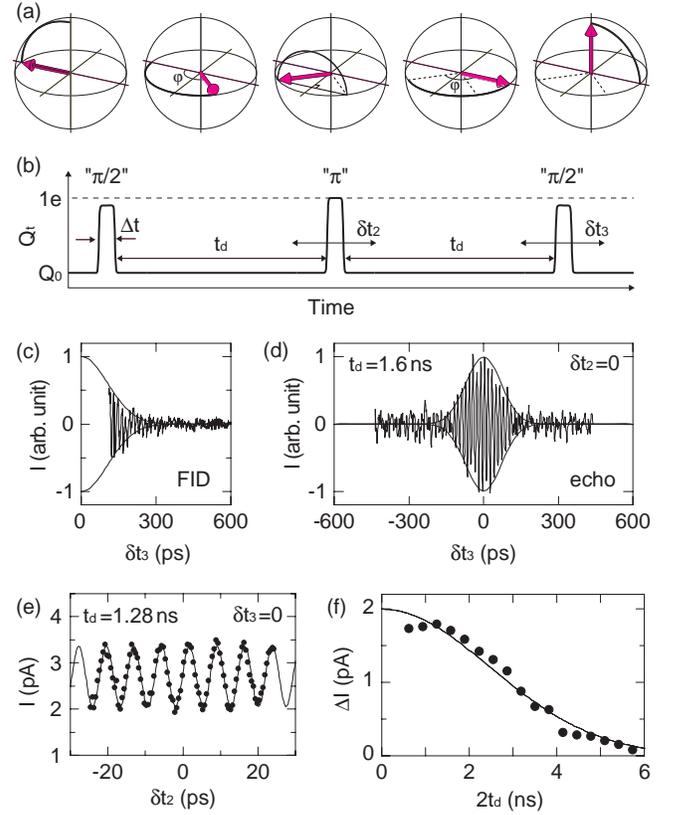}
\end{center}
\narrowtext
\caption{
Charge-echo experiment: 
(a) Schematic quantum-state evolutions. 
(b) Pulse sequence.
(c) Normalized free-induction decay (FID) signal vs.\ $\delta t_3$ taken without the second pulse and with $t_d =0$.
The oscillating signal is highpass-filtered and normalized to the gaussian envelope $\exp (-(\delta t_3/150 {\rm ps})^2)$.
(d) Normalized echo signal vs.\ $\delta t_3$. 
The envelope is $\exp (-(\delta t_3/100 {\rm ps})^2)$.
The signal-to-noise ratio is poor, because the data was taken with three ``$(\frac{\pi}{2})$''-pulses instead of the ideal pulse sequence.
$T_r$ is 64~ns in (c) and (d).
(e) Echo-signal current $I$ vs.\ $\delta t_2$.
Solid curve is a sinusoidal fit.
(f) Oscillation amplitude of the echo-signal current $\Delta I$ as a function of $2t_d$ with a gaussian fit.
}\label{fig:echo}
\end{figure}

\vspace{-5mm}

\begin{figure}
\begin{center}
\includegraphics[width=60mm]{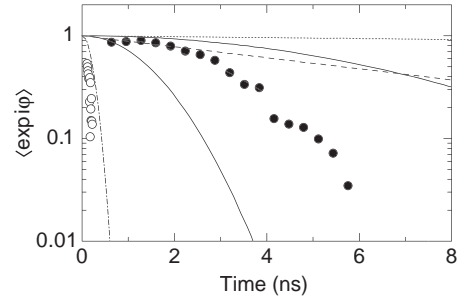}
\end{center}
\narrowtext
\caption{
Decay of the normalized amplitude of the echo signal (filled circles) and the FID signal (open circles) compared with estimated decoherence factors $\langle \exp i \varphi \rangle$ due to  electromagnetic environment (dotted line), the read-out process (dashed), and $1/f$ charge noise with $\alpha=(1.3 \times 10^{-3} e)^2$ (dash-dotted). 
Two solid lines are estimations for the echo experiment in the presence of the same $1/f$ charge-noise spectrum (bottom) and that with $\alpha=(3.0 \times 10^{-4} e)^2$ (top).
}\label{fig:decoherence}
\end{figure}

\end{multicols}


\begin{thebibliography}{99}
\bibitem{comment2}The Josephson junction is actually two parallel junctions forming a SQUID.
Throughout this work, the penetrating magnetic flux is fixed at $\phi/\phi_0=1/3$, where $\phi_0$ is a flux quanta, that gives the effective $E_J$.
\bibitem{comment3}Non-rectangular pulse shape and the finite ratio $\delta E/E_J$ modify the quantum evolution.
Additionally, jittering of pulses can reduce the signal size but do not change the decay time, either.
\end{thebibliography}
\end{document}